\documentclass[journal]{IEEEtran}
\usepackage{etoolbox}
\usepackage{cite}
\usepackage{graphicx}
\DeclareGraphicsExtensions{.pdf,.jpeg,.png}
\usepackage{amsmath}
\usepackage{algorithmic}
\usepackage{dsfont}
\usepackage{amsthm}
\usepackage{amssymb}
\usepackage{color}
\usepackage{array}
\usepackage{multirow}
\usepackage[caption=false,font=footnotesize]{subfig}
\newcommand{\calG}{\mathcal{G}}
\newcommand{\calV}{\mathcal{V}}
\newcommand{\calE}{\mathcal{E}}
\newcommand{\calX}{\mathcal{X}}
\newcommand{\eij}{e_{ij}}

\makeatletter
\patchcmd{\@maketitle}
  {\addvspace{0.5\baselineskip}\egroup}
  {\addvspace{-1\baselineskip}\egroup}
\makeatother

\begin{document}
\title{Chance-Constrained Optimal Distribution Network Partitioning to Enhance Grid Resilience}
\author{Shuchismita~Biswas,~\IEEEmembership{Student~Member,~IEEE,}
        Manish~K.~Singh,~\IEEEmembership{Student~Member,~IEEE,}
        and~Virgilio~A.~Centeno,~\IEEEmembership{Senior~Member,~IEEE}
\thanks{The authors are with Power and Energy Center (PEC), Department of Electrical and Computer Engineering, Virginia Tech, Blacksburg, VA, U.S.A. Emails : \{suchi,manishks,virgilio\}@vt.edu.}
}


\markboth{DRAFT VERSION}{Biswas, Singh, and Centeno: Chance-Constrained Optimal Distribution Network Partitioning to Enhance Grid Resilience}
\maketitle

\begin{abstract}
This paper formulates a chance-constrained optimal distribution network partitioning (ODNP) problem addressing uncertainties in load and renewable energy generation; and presents a solution methodology using sample average approximation (SAA). The objective is to identify potential sub-networks in the existing distribution grid; that are likely to survive as self-adequate islands if supply from the main grid is lost. {This constitutes a planning problem.} Practical constraints like ensuring network radiality and availability of grid-forming generators are considered. Quality of the obtained solution is evaluated by comparison with- a) an upper bound on the probability that the identified islands are supply-deficient, and b) a lower bound on the optimal value of the true problem.  Performance of the ODNP formulation is illustrated on a modified IEEE 37-bus feeder. It is shown that the network flexibility is well utilized; the partitioning changes with risk budget; and that the SAA method is able to yield good quality solutions with modest computation cost. 
\end{abstract}
\begin{IEEEkeywords}
microgrids, renewable energy generation, radiality, chance-constrained optimization, resilience, DER%
\vspace{-0.09in}
\end{IEEEkeywords}

\vspace{-0.1in}\section{Introduction}\label{sec: 1_intro}
\vspace{-0.04in}

\IEEEPARstart{I}n recent years, the adoption of renewable energy based distributed energy resources (DERs) has increased due to the recognition of their economic and environmental benefits. A primary advantage of DERs is their ability to sustain local loads if the main grid is lost, possibly due to natural disasters. DERs and loads may be clustered together to form \emph{microgrids}, a resiliency resource, which supply essential loads and aid service restoration during and after outages \cite{micro_resilience}. According to the IEEE 1547.4-2011 standard, microgrids: \emph{1)} have DERs and load; \emph{2)} can operate in both grid-connected and islanded modes;
and \emph{3)} are intentionally planned\cite{1547}. For safe operations, microgrids must have adequate control capabilities. 

Utilities are interested in identifying parts of the existing distribution network that can be converted to microgrids via economically viable retrofitting.
{ This is because DERs cannot supply local loads during an outage if adequate control and protection schemes are not in place.} Hence, optimally splitting a network into microgrids constitutes a pertinent { planning} problem~\cite{arxiv_radiality_19,TPWRS_17_Ding,TPWRS_15_wang,TSG_16_chen,TSG_19_barani,18_jovanovic_ACO,systems_19_osama,TSG_12_arefifar,TSG_16_Nasar}. This optimal distribution network partitioning (ODNP) task seeks to identify potential \emph{self-adequate} sub-networks that can survive the loss of the main grid as islands. Both exact \cite{arxiv_radiality_19,TPWRS_15_wang,TPWRS_17_Ding,TSG_19_barani,TSG_16_chen} and heuristic \cite{systems_19_osama,TSG_12_arefifar,18_jovanovic_ACO,TSG_16_Nasar} methods have been proposed for ODNP. Self-adequacy in the objective function has been surrogated by either expected power flow on microgrid boundary lines \cite{TSG_12_arefifar, systems_19_osama,TSG_19_barani,TSG_16_Nasar}, or expected load-generation imbalance within microgrids \cite{arxiv_radiality_19,TPWRS_15_wang,TPWRS_17_Ding,TSG_19_barani,TSG_16_chen,18_jovanovic_ACO}. 
Moreover, dynamic identification of boundary lines in response to faults have also been proposed \cite{TSG_16_chen,TSG_19_barani}.
A method for determining self-sufficient islands in transmission networks is described in \cite{biswas2019proactive}, but cannot be directly extended to ODNP without including distribution system specific constraints.

Distribution networks are usually operated radially for protection coordination, and this radiality needs to be maintained while separating into microgrids. In \cite{systems_19_osama,TPWRS_15_wang,TSG_12_arefifar,TSG_19_barani,TSG_16_Nasar,TSG_16_chen}, ODNP is demonstrated on an already radial feeder and radiality is not explicitly enforced. This approach ignores normally open switches, and under-utilizes network flexibility. Radiality is considered in \cite{TPWRS_17_Ding}, but another restrictive condition is imposed- each microgrid is assigned exactly one DER. This single DER constraint is also present in \cite{TSG_16_chen}. 
In this approach, the number of partitions are predetermined, leading to sub-optimal solutions. A radiality constraint without specifying the number of microgrids was recently presented in \cite{arxiv_radiality_19}, and the formulation in the current work builds upon this approach.
 
 A critical aspect that has been overlooked in the existing microgrid planning literature is the requirement of grid-forming generators in viable islands. The 1547 standard mandates that an island should have at least one generator that provides voltage and frequency support during a system disturbance, or has black-start capabilities \cite{1547}.  
 An exhaustive path search based method for checking connectivity to black-start generators has been proposed in \cite{Manish_GM_19}. Another multiple commodity flow based approach 
 outlined in \cite{bs} separately checks nodes for their connectivity to black-start nodes. Both these approaches become computationally prohibitive for large networks. 
 The ODNP formulation put forth in the present work guarantees that all nodes in each microgrid will be connected to at least one grid-forming generator. The formulation is somewhat similar to the single commodity flow model of \cite{bs} but uses fewer constraints and shows faster performance (empirically observed to be 10 to 20\% faster).
 
ODNP is further complicated by the uncertainty in demands and available generation capacity. In \cite{systems_19_osama,TPWRS_15_wang,TSG_12_arefifar,TSG_16_Nasar,TSG_19_barani}, the load-generation uncertainty is addressed by constructing typical daily profiles, over which optimization is performed. However, the quality of solution obtained is not evaluated. The present work formulates a chance-constrained ODNP (cc-ODNP) to identify optimal microgrids in the planning stage. This is computationally challenging as the underlying deterministic formulation uses mixed integer linear program (MILP) and is not convex by nature. Hence, sampling and integer programming \cite{cco} has been used to solve an approximation of the cc-ODNP, and the quality of the solution obtained is rigorously evaluated.  

The main contributions of this work are as follows. \emph{First}, a deterministic ODNP for identifying optimal microgrids, given real-time load-generation values is formulated in Section~\ref{sec: 2_prob}. {Practical constraints are comprehensively addressed, with the following novel aspects: }\emph{i)} efficiently formulating a generalized radiality condition, and \emph{ii)} ensuring every microgrid includes a grid-forming generator, without any pre-assignment. Thus, optimality is not compromised. \emph{Second,} a probabilistic ODNP problem is formulated, and solved using a computationally tractable sample average approximation (SAA) based MILP. While the SAA approach offers asymptotic equivalence to the original probabilistic formulation, in practice, computational resources restrict the number of scenarios that can be analyzed. Hence, rigorous stochastic tools have been used in section~\ref{sub-sec: validation} to statistically assess the quality of an obtained solution, in terms of confidence in feasibility and relevance of the attained objective. \emph{Third,} in section~\ref{sec: 3_results}, 
performance of ODNP is demonstrated through extensive numerical tests on a modified IEEE 37-bus feeder. It is shown that the SAA approach { is able to efficiently utilize network flexibility, and} outperforms a robust clustering based methodology in terms of objective cost.
\vspace{-0.1in}
\vspace{-0.1in}
\section{Preliminaries}\label{sec: 1a_prelim}
\vspace{-0.03in}

In this section, some mathematical preliminaries are revisited before expounding on the problem formulation. Calligraphic symbols represent sets, lower case bold letters represent column vectors, and upper case bold letters denote matrices. All zero and all one vectors and matrices of appropriate size are denoted by $\mathbf{0}$ and $\mathbf{1}$ respectively.
\vspace{-0.13in}
\subsection{Graph Theory}\label{sub_sec: graph theory}\vspace{-0.03in}
A graph  $\calG :=(\calV,\calE)$ consists of a \emph{vertex set} $\calV$ and an \emph{edge set} $\calE$, where an edge is an unordered pair of distinct vertices of $\calG$. Edge $e_{ij}\in\calE$ is denoted by its incident vertices $(i,j)$, such that $i,j\in\calV$. If $e_{ij}\in\calE$, then vertices $i$ and $j$ are \textit{adjacent}. 
Two edges are adjacent if they have a common vertex. A \textit{subgraph} of $\calG$ is a graph $\mathcal{H}:=(\mathcal{X,Y})$ such that $\mathcal{X}\subseteq\calV$ and $\mathcal{Y}\subseteq\calE$. If $\mathcal{X}=\calV$, then $\mathcal{H}$ is a \textit{spanning subgraph} of $\calG$. $\mathcal{H}$ is an \textit{induced subgraph} of $\calG$  if vertices in $\mathcal{X}$ are adjacent in $\mathcal{H}$ if and only if they are adjacent in $\calG$. 

A \textit{path} from $i$ to $j$ is a sequence of distinct vertices starting at $i$ and ending at $j$ such that consecutive vertices are adjacent. If there is a path between all pairs of vertices of a graph $\calG$, then $\calG$ is \textit{connected}; else $\calG$ is \textit{disconnected}. An
induced subgraph of $\calG$ that is maximal, subject to being connected, is called a \textit{connected component} of $\calG$. A \textit{cycle} is a sequence of adjacent edges without repetition that starts and ends at the same  node. A  graph  with  no  cycles  is called \textit{acyclic}.  A  connected and acyclic graph is a \textit{tree}. A \textit{spanning tree} subgraph of $\calG$ is a tree that covers all vertices in $\calG$. An acyclic graph with multiple connected components is a \textit{forest}. A \textit{spanning forest} subgraph of $\calG$ is a forest that covers all vertices in $\calG$. Spanning forests may include connected components with a single node. A review of graph theory fundamentals is available in \cite{Godsil2001}.
\vspace{-0.15in}
\subsection{Chance-Constrained Optimization}\label{sub_sec: cco} \vspace{-0.05in}
Stochastic optimization refers to a collection of methods for solving an optimization problem with uncertain parameters. 
For many real-world applications operating in uncertain environments, ensuring 100\% reliability is physically and economically impractical. This difficulty is often dealt with by designing systems that assure a minimum reliability level with high probability. Mathematical models of such reliability-constrained systems involve the use of probabilistic or \textit{chance constraints} \cite{cco}. A generic chance-constrained optimization (CCO) problem is of the form
\begin{align}\label{eq:cco}
\min_{\boldsymbol{x} \in \mathcal{X}} \quad &f(\boldsymbol{x}) \tag{$P_1$}\\
\text{s. to} \quad & \boldsymbol{h}(\boldsymbol{x}) \leq \boldsymbol {0}\tag{$C_1$}\\
\quad & Pr\{\boldsymbol{g}(\boldsymbol{x},\boldsymbol{\xi})\leq \boldsymbol{0}\}\geq 1-\varepsilon \tag{$C_2$} 
\end{align}

Here, $\boldsymbol{x}$ is the vector of decision variables, whose feasible region is given by $\mathcal{X}\subset \mathds{R}^n$. The objective function to be minimized is $f:\mathds{R}^n  \rightarrow \mathds{R}$. Vector $\boldsymbol{\xi}$ stacks the uncertain parameters with known probability distribution, and $\varepsilon \in (0,1)$ is a tunable risk parameter. Problem $P_1$ seeks to  find an optimal decision vector $\boldsymbol{x}^*$ that minimizes $f(\boldsymbol{x})$, such that the \textit{hard constraints} $C_1$ are always satisfied, while the \textit{chance constraint} $C_2$ is satisfied with probability at least $1-\varepsilon$.

In power systems literature, CCO   has   been   previously used  to  address  security  constrained  economic  dispatch  and unit  commitment  problems  \cite{cco_review}. This class of problems is difficult to solve, due to two main reasons:

\textbullet \quad Given a candidate solution $\boldsymbol{\Bar{x}}\in\calX$, accurately computing  $Pr\{\boldsymbol{g}(\boldsymbol{\Bar{x}},\boldsymbol{\xi})\leq \mathbf{0}\}$ can be very difficult, making it hard to check if constraint $C_2$ is satisfied.

\textbullet \quad The feasibility region defined by a chance constraint is usually not convex \cite{cco}. This makes finding an optimal solution difficult even when the feasibility of $\boldsymbol{\Bar{x}}$ can be checked.

These difficulties may be overcome by considering a \textit{sample average approximation} (SAA) of the original problem where the true distribution of $\boldsymbol{\xi}$ is replaced by an empirical distribution with discrete support. The SAA is still a chance-constrained stochastic problem, but with a different distribution for $\boldsymbol{\xi}$, and may be solved via integer programming \cite{cco}. This method has been shown to yield good candidate solutions if the sampling is ample and rich. In this work, the SAA approach will be incorporated to solve a probabilistic ODNP and the solution obtained will be further analyzed to verify how well it solves the original chance constrained problem.
\vspace{-0.15in}
\section{Problem Formulation}\label{sec: 2_prob}
\vspace{-0.05in}
Given a distribution network with DERs, planners would like to optimally construct microgrids, such that DERs sustain internal loads if supply from the main grid is lost. Load served is to be maximized. Both load and generation vary with weather and assuring self-adequacy for the worst case may lead to very conservative solutions. Hence, a solution that works well for \textit{most} operating conditions might be preferred. Thus microgrids may be designed to be self-adequate with probability at least $(1-\varepsilon)$ across all possible operating scenarios, where $\varepsilon$ is a tunable risk parameter. The value of $\varepsilon$ may be chosen based on available storage resources. Once optimal microgrids are identified, they need to be equipped with control capabilities and boundary line switches. It must be noted that depending on the generation capacity  of installed DERs, all load may not be served by microgrids.

Our mathematical formulation is put forth in three steps. First, a deterministic version of the problem, d-ODNP is presented where the load served is maximized for a given scenario of demands and generation. Next, the chance constraints arising from the randomness in generation and demands are added. DERs are assumed to be dispatchable subject to stochastic generation capacity. Such units in practice could be photo-voltaic (PV) generators, diesel generators (DGs) or combined heat and power plants (CHPs) that are plausible in a low/medium voltage network setup. Non-dispatchable generators may be seamlessly incorporated in the formulation as negative stochastic demands. Finally, a SAA based algorithm is proposed that can tractably solve the probabilistic ODNP.
\vspace{-0.15in}
\subsection{Distribution Network Model}
\vspace{-0.03in}
A single-phase distribution network may be represented by a connected directed graph $\calG_N:=(\calV_N,\calE_N)$, where vertices denote buses and edges denote lines. The substation node is indexed by $0$; and the set of all other nodes is denoted by $\calV:=\calV_N\setminus\{0\}$. Each edge $e_{i,j}\in\calE_N$ is assigned an arbitrary direction from node $i$ to $j$. If $e_{i,j}\in\calE_N$, then $e_{j,i}\notin\calE_N$. The task at hand considers that the main grid is unavailable, hence partitioning needs to be carried out on $\calG:=(\calV,\calE)$, the induced subgraph of $\calG_N$ on vertex set $\calV$. In the present setup all lines are considered switchable. Any non-switchable edge coinciding with a microgrid boundary would need to be retrofitted with a switch. Moreover, edges include lines with existing normally open and normally closed switches, and hence $\calG$ is not necessarily radial. 

Each node has an associated demand $(\xi^{d_p}_i+j\xi^{d_q}_i)$ and generation capacity $(\xi^{g_p}_i+j\xi^{g_q}_i)$. The demand and generation capacities are not precisely known at the planning stage and only a probability distribution, possibly empirical, may be available. Let $v_i$ be the  voltage magnitude at bus $i$ and $(p_i+jq_i)$ be the complex power injection. Bus voltages, demand, generation capacity, and complex power injections are respectively stacked into vectors $\mathbf{v}$, $\boldsymbol{\xi^{d_p}}+j\boldsymbol{\xi^{d_p}}, \boldsymbol{\xi^{g_p}}+j\boldsymbol{\xi^{g_q}},\mathbf{p}+j\mathbf{q}$.  All quantities are in per units.

Let us introduce two binary decision variables $b^n_i\in\{0,1\}$ and $b^e_{ij}\in\{0,1\}$ that respectively dictate if vertex $i\in\calV$ and edge $e_{ij}\in\calE$ are energized. If $e_{ij}$ is energized, then adjacent vertices $i$ and $j$ need to be energized as well. Mathematically,
\begin{equation}
     b^n_i+b^n_j \geq 2b^e_{ij} \quad \forall { \eij \in \calE}\label{constr:topo}
\end{equation}
For safe operations, ANSI standards mandate that voltages at active buses should be within $\pm 5\%$ p.u. of the nominal value \cite{ANSI}. Mathematically,
\begin{equation}
0.95~\mathbf{b^n} \leq \mathbf{v} \leq 1.05~\mathbf{b^n}.\label{constr:v}
\end{equation}
Let the power flow on line $\eij \in \calE$ be $P_{ij}+jQ_{ij}$. The line capacity constraints may be formulated as follows. 
\begin{align}
    -b^e_{ij}P^{max}_{ij} \leq P_{ij} \leq b^e_{ij}P^{max}_{ij} \quad \forall{e_{ij} \in \calE}\tag{3a}\\
    -b^e_{ij}Q^{max}_{ij} \leq Q_{ij} \leq b^e_{ij}Q^{max}_{ij} \quad \forall{e_{ij} \in \calE}\tag{3b}
\end{align}
Flow constraints of the form $P_{ij}^2+Q_{ij}^2\leq S_{ij}^2$ are not used here to avoid quadratic constraints. A polytopic approximation of this constraint proposed in \cite{apparent_jabr} could also be used.
\vspace{-0.18in}
\subsection{Power Flow Model}\vspace{-0.05in}
Power injection at energized buses is assumed to be controllable subject to the maximum capacity, while loads are assumed to be inelastic. { Some generators can absorb reactive power, and this flexibility is included in the formulation as negative reactive power generation. Let the maximum reactive power absorption at bus $i$ be denoted as $q_i^{min}$. The value of $q_i^{min}$ can be set to zero to indicate the absence of reactive power absorption capabilities.} The nodal power injections are governed by the following expressions:
\begin{align}
    0\leq p^g_i \leq b^n_i \xi^{g_p}_i \quad \forall {i \in \calV} \label{eq:pgen}\tag{4a}\\
    -b^n_iq_i^{min}\leq q^g_i \leq b^n_i \xi^{g_q}_i \quad \forall {i \in \calV} \label{eq:qgen}\tag{4b}
    \end{align}
Here, constraints \eqref{eq:pgen}-\eqref{eq:qgen} establish that generation output $p^g_i,q^g_i$ at bus $i$ is {bounded}. The maximum capacity of DGs may be fixed based on the machine rating. The active power capacity of PV generators is limited by solar irradiance levels. Similarly, the maximum active power generation by CHPs is affected by the local heating demand. These factors make $\xi^{g_p}$ stochastic, in general. Furthermore, given the limit on active power, a corresponding stochastic limit on reactive power may be obtained based on the operating curves of the respective generators. For inelastic loads, the constraints for power consumption may be formulated as:
    \begin{align}
    p^d_i = b^n_i \xi^{d_p}_i \quad \forall {i \in \calV} \label{eq:pload}\tag{5a}\\
    q^d_i = b^n_i \xi^{d_q}_i \quad \forall {i \in \calV}\label{eq:qload}\tag{5b}
    \end{align}
    Constraints \eqref{eq:pload}-\eqref{eq:qload} state that consumption at bus $i$ is equal to its demand; if energized. Thus the net power injections are:
    \begin{align}
    p_i=p^g_i-p^d_i \quad \forall {i \in \calV} \label{eq:pinject}\tag{6a}\\
    q_i=q^g_i-q^d_i \quad \forall {i \in \calV} \label{eq:qinject}\tag{6b}
    \end{align}
For the power flow, the linearized distribution flow (LDF) model proposed in \cite{LDF} is followed. Despite being an approximation for the full AC power flow model, LDF has been used extensively and shown to perform well in literature~\cite{LDF_ex}. Thus, ignoring line losses, the power balance at each node entails:
\begin{align}
    \sum_{e_{ij}\in\calE} P_{ij} -  \sum_{e_{jk}\in\calE} P_{jk} = p_j \quad \forall{j\in\calV}\tag{7a}\\
    \sum_{e_{ij}\in\calE} Q_{ij} -  \sum_{e_{jk}\in\calE} Q_{jk} = q_j \quad \forall{j\in\calV}\tag{7b}
\end{align}
Let $r_{ij}+jx_{ij}$ be the impedance of line $e_{ij}\in\calE$. Then, the relationship between voltages and power injections may be linearized as: $v^2_i-v^2_j=2(r_{ij}P_{ij}+x_{ij}Q{ij})$. Assuming small voltage deviations, the squared terms may be approximated as $v^2_i \simeq 2v_i - 1$. Using these results, 
\begin{align}
    b^e_{ij}(v_i-v_j-r_{ij}P_{ij}-x_{ij}Q_{ij})=0,~\forall e_{ij}\in\calE \label{LDF}\tag{8}
\end{align}
Here, the indicator $b^e_{ij}$ is multiplied to enforce the voltage drop relation only for the energized lines. Bilinear terms like $b^e_{ij}v_i$ in \eqref{LDF} can be handled by McCormick linearization, where the  product terms are replaced by  their  linear  convex  envelopes  to  yield  a  relaxation  of  the original  non-convex  feasible  set \cite{McCormick1976}.  If  at  most  one  of  the  variables  is  continuous  and  the  rest  are  binary,  this relaxation is exact.  For illustration, let us consider a term $z=xy$, where $x$ is binary and $y$ is a continuous variable bounded in $y\in[\underline{y},\overline{y}]$. Here, $z=xy$ may be equivalently expressed as four linear inequality constraints.
\begin{align}
    x\underline{y}&\leq z\leq x\overline{y}\label{eq:mccormick1}\tag{9a}\\
    y+(x-1)\overline{y}&\leq z\leq y + (x-1)\underline{y}\label{eq:mccormick2}\tag{9b}
\end{align}
Note that putting  $x=0$ in \eqref{eq:mccormick1}-\eqref{eq:mccormick2} yields $z=0$. Similarly, putting $x=1$ yields $z=y$. All such bilinear terms appearing henceforth in this work will be treated in a similar manner.

\subsection{Radiality Constraint}\vspace{-0.03in}
Network radiality is essential for distribution system operations. Some approaches proposed for enforcing radiality are cycle elimination \cite{Manish_GM_19} and virtual commodity flow \cite{arxiv_radiality_19,Manish_PSCC_20}. The ODNP task needs to identify $\calG'=(\calV,\calE')$, a spanning subgraph of $\calG$, such that every connected component, or simply component, of $\calG'$ is a tree, i.e. $\calG'$ is a \textit{forest}. A spanning forest may include isolated nodes, i.e. it may have components with a single node. 
The radiality constraints in this context will be formulated using the condition stated next.

\begin{figure}
    \centering
    \includegraphics[width=0.48\columnwidth]{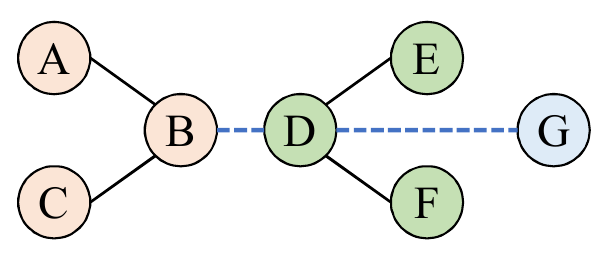}
    \caption{Components $(\{A,B,C\},\{AB,BC\})$,$(\{D,E,F\},\{DE,DF\})$ and $(\{G\},\{\})$ form a spanning forest. Adding edges $BD$ and $DG$ creates a spanning tree. }
    \label{fig:forest}
    \vspace{-0.27in}
\end{figure}

\textbf{Proposition 1.}(\hspace{-0.1pt}\cite{arxiv_radiality_19})~\textit{Given a spanning forest subgraph $\mathcal{F}:=(\calV,\calE_F)$ of a connected graph $\calG:=(\calV,\calE)$, there exists at least one spanning tree subgraph of $\calG$, expressed as $\mathcal{T}:=(\calV,\calE_T)$, such that $\calE_F \subseteq \calE_T \subseteq \calE$.}

In other words, some edges may be removed from a spanning tree to obtain a spanning forest. This idea is illustrated in fig. \ref{fig:forest}. The solid lines show edges in a spanning forest and the addition of dashed edges creates a spanning tree. Hence, the radiality of $\calG':=(\calV,\calE')$ holds true if there exists a fictitious spanning tree subgraph $\mathcal{T}:=(\calV,\calE_T)$ of $\calG$ such that $\calE'\subseteq\calE_T$. 
Let us first establish a condition to select a spanning tree, and then extract the required spanning forest from it. The base topology of $\calG$ may be captured by a branch-bus incidence matrix $\mathbf{\Tilde{A}}$ of dimension $|\calE| \times (|\calV|)$, with the following entries:
\begin{equation}
    \mathbf{\Tilde{A}}_{e_{ij},k}:= 
\begin{cases}
    1\quad &,\quad k=i\\
    -1\quad &,\quad k=j\\
   0              &, \quad \text{otherwise}
\end{cases}\label{eq:A}\tag{10}
\end{equation}
The first column $\mathbf{a}_1$ of $\mathbf{\Tilde{A}}$ may be  separated as $\mathbf{\Tilde{A}}=[\mathbf{a}_1 \hspace{3pt} \mathbf{A}]$.  This yields the \textit{reduced} branch-bus incidence matrix $\mathbf{A}$ of $\calG$. An efficient model for imposing graph connectivity put forth in \cite{Manish_PSCC_20} posits that a graph with vertex set $\calV$ and reduced branch-bus incidence matrix $\mathbf{A}$, is connected if and only if there exists a vector $\mathbf{f}\in \mathds{R}^{|\calE|}$, such that $\mathbf{A}^T\mathbf{f}=\mathbf{1}$. For proof, see \cite{Manish_PSCC_20}. For a physical interpretation, consider every vertex in $\calV\setminus\{1\}$ injects unit virtual commodity into the network represented by the graph. Then, $\mathbf{f}$ denotes the flow of commodities on the edges. If this flow setup is feasible, then there must be a withdrawal of $|\calV|-1$ units at vertex $1$, and every vertex in $\calV\setminus\{1\}$ must have a path to reach vertex $1$. Vertex $1$ may be arbitrarily chosen.
Stating a well-known result from graph theory, a tree with $n$ vertices has exactly $n-1$ edges. Hence, the radiality constraints become:
\begin{align}
    \mathbf{A}^T\mathbf{f}&=\mathbf{1} \label{eq:connectivity}\tag{11a}\\
    -(|\calV|-1)\boldsymbol{\theta}\leq & \text{ }\mathbf{f}\leq(|\calV|-1)\boldsymbol{\theta},\label{eq:flow}\tag{11b}\\
    \boldsymbol{\theta}&\in\{0,1\}^{|\calE|} \tag{11c}\\
    \mathbf{1}^T\boldsymbol{\theta}&=|\calV|-1\label{eq:tree}\tag{11d}\\
    \mathbf{b^e}&\leq\boldsymbol{\theta}\label{eq:forest}\tag{11e}
\end{align}
Constraints \eqref{eq:connectivity}-\eqref{eq:tree} ensure that auxiliary binary indicator variables $\boldsymbol{\theta}$ on the edge-set of the base graph describe a spanning tree. Then, \eqref{eq:forest} states that the edges selected via the binary variables $\mathbf{b^e}$ are a subset of this spanning tree; and hence form a spanning forest structure as per Proposition~1. Since the radiality constraints are thus posed, the number of components in $\calG'$ 
need not be pre-assigned. 
\vspace{-0.15in}
\subsection{Connection to grid-forming generators}
As stated in Section \ref{sec: 1_intro}, each microgrid should have a grid-forming generator. 
Let $\calV_{s}$ be the set of buses with grid-forming capabilities.

Then, the connectivity constraint becomes:
\begin{align}
    \sum_{e_{jk}\in\calE} f'_{jk}-\sum_{e_{ij}\in\calE} f'_{ij} &=b^n_j, \forall j\in\calV\setminus\calV_{s}\label{eq:flow_bs}\tag{12a}\\
    -(|\calV|-|\calV_{s}|)\mathbf{b^e} \leq \mathbf{f'} &\leq (|\calV|-|\calV_{s}|)\mathbf{b^e}\label{eq:flowbound}\tag{12b}
\end{align}
Constraint \eqref{eq:flow_bs} states that every energized non grid-forming node in $\calG$ injects unit virtual commodity into the network. 
 Constraint \eqref{eq:flowbound} bounds flows on energized lines and fixes flows on deenergized ones at 0. Here, $\mathbf{f'}\in\mathds{R}^{|\calE|}$ is a vector representing virtual line flows. It must be emphasized that $\mathbf{f'}$ is different from $\mathbf{f}$ in \eqref{eq:connectivity}. Both these vectors are used to impose connectivity conditions, and have no physical significance related to the actual power flow. Again, this setup is feasible only when all units injected by energized non grid-forming nodes can be withdrawn at grid-forming nodes. Some grid-forming nodes may not be energized in the optimal topology. This is implicitly considered through the topological constraint in \eqref{constr:topo}, that ensures all edges connected to a deenergized node are deenergized as well. Therefore, no path exists from an energized non grid-forming to a deenergized grid-forming node.   
 {If multiple buses in a microgrid host generators with grid-forming capabilities, only one must be assigned as the reference bus that determines the system operating point. Power sharing strategies among multiple dispatchable generators in a microgrid have been widely studied in literature, see \cite{powersharing2} and references therein.}
  
\vspace{-0.15in}
\subsection{Deterministic ODNP}\label{sub-sec: deterministic prob}

The  d-ODNP is solved for one realisation of the power generation capacity $(\boldsymbol{\xi^{g_p}},\boldsymbol{\xi^{g_q}})$, and demands  $(\boldsymbol{\xi^{d_p}}, \boldsymbol{\xi^{d_q}})$. The central idea is to sustain maximum load through microgrids if supply from the main grid is lost, thereby minimizing service interruption. Therefore, the objective for d-ODNP becomes maximizing load served. The entire deterministic optimization setup may be mathematically expressed as follows. 
\begin{align}
    \min \quad &{-\mathbf{1}^T\mathbf{p^d}}\label{ODNP-1}\tag{ODNP-1}\\
    \text{s. to} \quad &\eqref{constr:topo}-\eqref{LDF},\eqref{eq:connectivity}-\eqref{eq:forest},\eqref{eq:flow_bs}-\eqref{eq:flowbound}\notag
\end{align}
The relative priority of loads has not been considered in (ODNP-1). However, this cost may be modified by assigning weights to loads in proportion to their criticality.
\vspace{-0.15in}
\subsection{Probabilistic ODNP}\label{sub-sec: chance}
The problem \eqref{ODNP-1} applies to one realization of the generation-demand scenario $(\boldsymbol{\xi^{g_p}},\boldsymbol{\xi^{g_q}},\boldsymbol{\xi^{d_p}}, \boldsymbol{\xi^{d_q}})$. However, a more realistic goal would be to identified microgrids that are optimal in some sense for a large set of realizations of the generation-demand scenarios. In the latter setup, the decision variables $\boldsymbol{\psi}_1:=\{\mathbf{b^n,~b^e,~f,~f'}\boldsymbol{,\theta}\}$ shall remain fixed for all realizations of the uncertainties. The realization dependent variables would be $\boldsymbol{\psi}_2:=\{\mathbf{p^d,~q^d,~p^g,~q^g,~v,~P,~Q}\}$. Collecting all the uncertainties in $\boldsymbol{\xi}:=\{\boldsymbol{\xi^{g_p}},\boldsymbol{\xi^{g_q}},\boldsymbol{\xi^{d_p}}, \boldsymbol{\xi^{d_q}}\}$, the probabilistic ODNP may seek to solve-
\begin{align}
    \min_{\boldsymbol{\psi_1}}~~~&-\mathds{E}_{\boldsymbol{\xi}} [\mathbf{1}^T\mathbf{p^d}]\label{ODNP-2}\tag{ODNP-2}\\
    \text{s. to}~~~&\textrm{Pr}\left(\exists~\boldsymbol{\psi_2}| \mathds{1}(\eqref{constr:topo}-\eqref{LDF},\eqref{eq:connectivity}-\eqref{eq:flowbound})=1\right)\geq 1-\epsilon\notag
\end{align}
The probabilistic constraint in \eqref{ODNP-2} is very difficult to enforce in practice. However, we will next discuss some reformulations that simplify the setup without loss of generality. First, note that if the power demands at all nodes are zero, then for a feasible $\boldsymbol{\psi}_1$ satisfying \eqref{constr:topo} and \eqref{eq:connectivity}-\eqref{eq:flowbound}, there always exist a $\boldsymbol{\psi}_2$ that satisfy all other constraints. Therefore, the probabilistic constraint may be equivalently posed by enforcing all constraints other than \eqref{eq:pload}-\eqref{eq:qload} as hard constraints, and putting the probabilistic requirement on \eqref{eq:pload}-\eqref{eq:qload}. Setting aside \eqref{eq:qload} for expository convenience, notice that the equality constraints in 
\eqref{eq:pload} may be decomposed into the following inequality constraints. 
\begin{align}
p^d_i-b^n_i\xi_i^{d_p}\leq 0\label{eq:p_hard},\quad \forall i\tag{14a}\\
-p^d_i+b^n_i\xi_i^{d_p}\leq 0 \label{eq:p_chance},\quad \forall i\tag{14b}
\end{align}
Now, \eqref{eq:p_hard} can be posed as a hard constraint, leaving \eqref{eq:p_chance} as the main chance constraint. To reiterate, the ODNP task seeks to identify potential microgrids within an existing distribution network, such that load served is maximized, and the microgrids are self-adequate with probability at least $(1-\varepsilon)$, if the main grid is lost.
Islanded  microgrids are called self-adequate when their internal load can be met by their internal generation.   Mathematically, 
\begin{align}
    Pr(-p^d_i+b^n_i\xi_i^{d_p}\leq0, \forall i )\geq 1-\varepsilon\label{eq:cc1} \tag{15}
\end{align}
The self-adequacy condition of microgrids may be thus posed at the node level since constraints \eqref{eq:connectivity}-\eqref{eq:forest} ensure that the network topology is a spanning forest, and hence loads within a microgrid can be supplied only from generators within the same microgrid. If needed, one may relax the the reliability requirement by modifying \eqref{eq:cc1} slightly. For instance, the condition, a microgrid should be able to meet 90\% of its internal load could be written as $Pr(-p^d_i+0.9\times b^n_i\xi_i^{d_p}\leq0, \forall i)\geq 1-\varepsilon\label{cc1}$.
\vspace{-0.25in}
\subsection{Sample Average Approximation}\label{sub-sec: SAA}
Recall from Section \ref{sub_sec: cco} that a chance constraint 
needs to be satisfied with a probability specified by a risk parameter $\varepsilon$. The chance-constraint $(C_2)$ may also be rewritten as $q( \boldsymbol{x}) \leq \varepsilon$, where $q(\boldsymbol{x})=Pr(\boldsymbol{g}(\boldsymbol{x},\boldsymbol{\xi})>\boldsymbol{0})$. Let $\boldsymbol{\xi^1},\boldsymbol{\xi^2},..., \boldsymbol{\xi^N}$ be $N$ independent and identically distributed (iid) samples of the uncertainty vector $\boldsymbol{\xi}$; then $\hat{q_N}(\boldsymbol{x})$, an estimator of $q(\boldsymbol{x})$ is equal to the proportion of realizations in the sample where $\boldsymbol{g}(\boldsymbol{x},\boldsymbol{\xi^i})>\boldsymbol{0}, i=1,..,N$. This is a sample average approximation of the chance-constrained problem ($P_1$) for the samples $\boldsymbol{\xi^1},\boldsymbol{\xi^2},\dots, \boldsymbol{\xi^N}$:
\[\min_{\boldsymbol{x} \in \mathcal{X}} \quad   f(\boldsymbol{x}) \quad \textrm{s.to.} \quad 
\hat{q_N}(\boldsymbol{x})\leq \gamma \tag{$P_2$} \]
Here, $\gamma\in(0,1)$ and is the risk level for the SAA problem. Assuming that the SAA can be solved, a) if $\gamma<\varepsilon,$ and $N$ is sufficiently large, SAA is a restriction on the true problem and a feasible solution of SAA is likely to be feasible for the true problem as well, b) if $\gamma>\varepsilon$, SAA is a relaxation of the true problem and the optimal value of SAA is likely to be a lower bound to the optimum for true problem. It can be shown that for $\gamma=\varepsilon$, the SAA optimum approaches its true counterpart with probability one as $N$ approaches infinity \cite{cco}. 

The chance-constrained SAA problem $P_2$ can be solved using MILP for $N$ iid samples of $\boldsymbol{\xi}$ as shown below\cite{cco}.
\begin{align}
\min_{\boldsymbol{x} \in \mathcal{X}} \quad &f(\boldsymbol{x}) \tag{$P_3$}\\
\text{s. to} \quad & \boldsymbol{h}(\boldsymbol{x}) \leq \boldsymbol{0} \tag{16a}\\
& \boldsymbol{g}(\boldsymbol{x},\boldsymbol{\xi}^\alpha)\leq M(1-z_\alpha), \quad & \alpha=1,2,\dots,N \label{eq:bigM}\tag{16b}\\
&\mathbf{1}^T\mathbf{z} \geq (1-\gamma) N\label{eq:cardinality}\tag{16c}\\
& \mathbf{z} \in \{0,1\}^N, \quad & \alpha=1,2,\dots,N\label{eq:saa}\tag{16d}
\end{align}
Here, $\alpha$ is used to index samples of $\boldsymbol{\xi}$, $z_\alpha$ is a binary variable and $M$ is a  large number such that $M>\max_{\boldsymbol{x} \in \mathcal{X}}\boldsymbol{g}(\boldsymbol{x},\boldsymbol{\xi}^\alpha)$ for all $\alpha=1,2,\dots,N$. Vector $\mathbf{z}$ stacks all $z_\alpha$ values. In constraint \eqref{eq:bigM}, if $z_\alpha$=1, then the chance constraint is not violated. If $z_\alpha=0$, then no bound is imposed. The cardinality constraint in \eqref{eq:cardinality} bounds the proportion of constraint violations. 


For ODNP the hard constraints are given by $\{\eqref{constr:topo}-\eqref{eq:qgen},\eqref{eq:qload}-\eqref{LDF},\eqref{eq:connectivity}-\eqref{eq:flowbound},\eqref{eq:p_hard} \hspace{2pt}\forall \alpha\}$. Probability of violating the chance-constraint $\{\eqref{eq:p_chance}\hspace{2pt}\forall \alpha\}$ is to be bounded.  Equation \eqref{eq:bigM} becomes:
\begin{align}
   -\mathbf{p^d}_\alpha+diag(\mathbf{b^n})\times\boldsymbol{\xi}^{\mathbf{d_p},\alpha}\leq M(1-z_\alpha)\times\mathbf{1},\quad \forall \alpha \label{eq:G}\tag{17}
\end{align}
Putting everything together, the problem becomes:
\begin{align}
    \min\quad &-\frac{1}{N}\sum_{\alpha=1}^N\mathbf{1}^T\mathbf{p^d}_\alpha\label{eq:ODNP3}\tag{ODNP-3}\\
    \text{s. to} \quad &\mathbf{p^d}_\alpha\leq z_\alpha \times \boldsymbol{\xi^{d_p}}, \quad \forall \alpha\label{eq:z}\tag{18}\\ &\eqref{constr:topo}-\eqref{eq:qgen},\eqref{eq:connectivity}-\eqref{eq:flowbound},\notag\eqref{eq:p_hard},\eqref{eq:cardinality}-\eqref{eq:G}\notag
\end{align}

The objective function in \eqref{eq:ODNP3} is the sample-based estimator of the objective in \eqref{ODNP-2} designed to maximize average load served across considered scenarios. Constraint \eqref{eq:z} fixes bus consumptions at zero when constraint \eqref{eq:p_chance} is not satisfied. This motivates the optimal solution for the ODNP to be one that also increases $\mathbf{1}^T\mathbf{z}$, lower bounded by $(1-\gamma)N$.  


{The optimal topology obtained by solving (ODNP-3) is determined by vectors $\mathbf{b^n}^*$ and $\mathbf{b^e}^*$. In practice, only lines connecting an energized node to a deenergized node will need to be disconnected to isolate the microgrids.}

\vspace{-0.13in}
\section{Solution Validation}\label{sub-sec: validation}
Consider a candidate solution $\boldsymbol{\Bar{x}}$ found by the SAA approach of \eqref{eq:ODNP3}.  To adjudge its quality, two aspects need to be analyzed: \emph{a)} Can it be said with some desired confidence that $\boldsymbol{\Bar{x}}$ a feasible solution for the true problem \eqref{ODNP-2}? \emph{b)} If yes, then how far is  $f(\boldsymbol{\Bar{x}})$ from the optimal value $f(\boldsymbol{x}^*)$? A method for checking 
an upper bound of $Pr\{\boldsymbol{g}(\boldsymbol{\Bar{x}},\boldsymbol{\xi})> \mathbf{0}\}$ and lower bound on $f(\boldsymbol{x}^*)$ is shown in \cite{cco} and references therein.

\textbullet\textit{ Upper bound on violation probability :}
Consider $N'$ iid realizations of $\boldsymbol{\xi}$, such that $N'>>N$, where $N$ is the number of $\boldsymbol{\xi}$ samples considered for solving the SAA problem. 
Here, $N'$ may be large as the samples will not be used in solving an optimization problem and hence not pose computational issues. Let $\hat{q}_{N'}(\boldsymbol{\Bar{x}})$ be an estimator of $q(\boldsymbol{\Bar{x}})$; equal to the proportion of times the event $\mathds{1}(\boldsymbol{g}(\boldsymbol{\bar{x}},\boldsymbol{\xi^j})>\boldsymbol{0})=1$ is observed in $N'$ trials. Estimator $\hat{q}_{N'}(\boldsymbol{\Bar{x}})$ of $q(\boldsymbol{\Bar{x}})$ is unbiased, implying $\mathds{E}(\hat{q}_{N'}(\boldsymbol{\Bar{x}}))=q(\boldsymbol{\Bar{x}})$. Also, for a large $N'$, its distribution may be approximated by a normal distribution with mean $q(\boldsymbol{\Bar{x}})$ and variance $q(\boldsymbol{\Bar{x}})(1-q(\boldsymbol{\Bar{x}}))/N'$ \cite{cco}. This yields an approximate $(1-\beta)$-confidence upper bound on $q(\boldsymbol{\Bar{x}})$:
\[
U_{\beta,N'}(\boldsymbol{\Bar{x}}):=\hat{q}_{N'}(\boldsymbol{\Bar{x}})+z_\beta\sqrt{\hat{q}_{N'}(\boldsymbol{\Bar{x}})(1-\hat{q}_{N'}(\boldsymbol{\Bar{x}}))/N'}
\]

Here, $z_\beta=\Phi^{-1}(1-\beta)$, where $\Phi$ is the  cumulative distribution function for the standard normal distribution, $\beta\in(0,1)$. We compare $U_{\beta,N'}(\boldsymbol{\Bar{x}})$ to $\varepsilon$ to check if $\boldsymbol{\bar{x}}$ is a feasible solution. 

\textbullet\textit{ Lower bound on optimal value :}
A procedure for deriving a lower bound for $f(\boldsymbol{x}^*)$ is shown in \cite{cco}. Let the SAA problem \eqref{eq:ODNP3} be solved for $N''$ iid samples of $\boldsymbol{\xi}$ and risk level $\gamma\geq 0$; and denote this problem by $P^{N''}_\gamma$. Let the true problem \eqref{ODNP-2} with risk $\varepsilon$ be denoted as $P_\varepsilon$. Now, the probability that at most $\lfloor\gamma N''\rfloor$ constraint violations are observed in $N''$ trials while solving $P^{N''}_\gamma$, when the true violation probability is $\varepsilon$, becomes:
\[\Theta_{N''} := B(\lfloor \gamma N'' \rfloor;\varepsilon,N'')\]
where,
\vspace{-2pt}
\[
B(k;q,N):=\sum_{r=0}^{k-1} \binom{N}{r}q^r(1-q)^{N-r}
\]
is the cumulative density function of the binomial distribution. 
Say, solving $P^{N''}_\gamma$ yields an objective value $f(\boldsymbol{\bar{x}})$. Assuming $P_\varepsilon$ has an optimal solution $f(\boldsymbol{x}^*)$, $Pr\{f(\boldsymbol{\bar{x}})\leq f(\boldsymbol{x}^*)\}\geq\Theta_N''$. 
This result yields a method for obtaining lower bounds with a specified confidence level $(1-\beta)$. Consider two positive integers $M$ and $N''$, such that $M>N''$. Generate $M$ independent sets of $N''$ iid samples of $\boldsymbol{\xi}$, and solve the SAA problem for each of the $M$ sets to obtain values $f(\boldsymbol{\bar{x}}_j),j=1,2,\dots,M$. These can be viewed as iid samples of the random variable $f(\boldsymbol{\bar{x}}))$. Let $L$ be the largest integer such that $B(L-1;\Theta_N'',M)\leq\beta$. If the optimal values  are arranged in a non-decreasing order $f(\boldsymbol{\bar{x}}_1)\leq f(\boldsymbol{\bar{x}}_2)\leq\dots f(\boldsymbol{\bar{x}}_M)$, it can be shown that with probability at least $(1-\beta)$, $f(\boldsymbol{\bar{x}}_L)$ is lower than the true optimal value $f(\boldsymbol{x^*})$.

Note that $f(\boldsymbol{\bar{x}}_L)>f(\boldsymbol{x^*})$ if and only if more than $L$ of the observed $f(\boldsymbol{\bar{x}}_j)$ values are greater than $f(\boldsymbol{x^*})$. Considering event $f(\boldsymbol{\bar{x}}_j)\leq f(\boldsymbol{x^*})$ as a success, $f(\boldsymbol{\bar{x}}_L)>f(\boldsymbol{x^*})$ if and only if there are fewer than $L$ successes in $M$ trials, with success probability $\Theta_{N''}$. Probability of fewer than $L$ successes in $M$ trials is $B(L-1,\Theta_{N''},M)$, and the bounding procedure described in this section restricts this probability value to $\beta$.

\vspace{-0.1in}
\section{Numerical Results}\label{sec: 3_results}
Performance of the proposed methodology is illustrated through computational experiments on a 3.6 GHz Intel Core i7-4790 CPU with 32 GB RAM. Optimization tasks are solved using YALMIP and Gurobi \cite{yalmip,gurobi}.
\vspace{-0.13in}
\subsection{Experiment Set-up}
\begin{figure}
    \centering
    \includegraphics[width=\columnwidth]{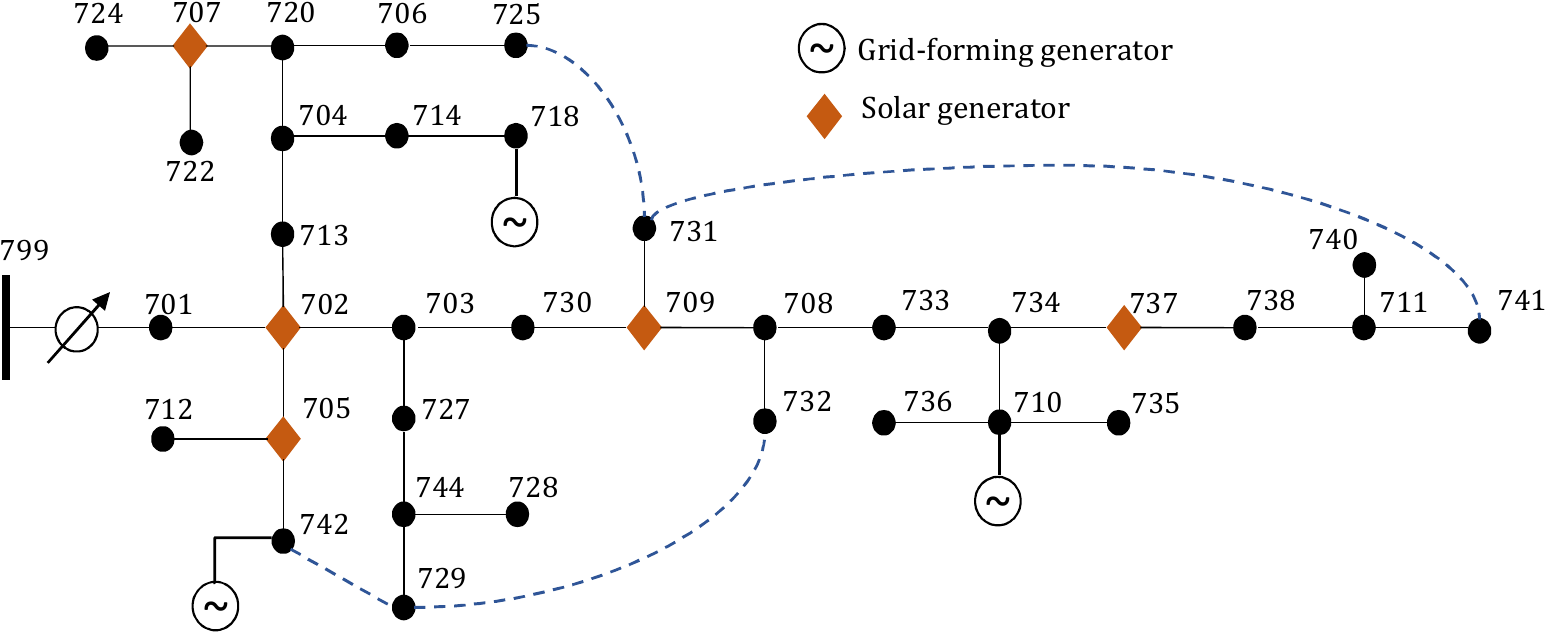}
    \caption{Modified IEEE 37-bus benchmark feeder with DER locations and normally open switches}
    \label{fig:37bus}
    \vspace{-0.22in}
\end{figure}
The ODNP problem is solved for a modified version of the IEEE 37-bus benchmark feeder (fig. \ref{fig:37bus}), converted to its single-phase equivalent by: a) assigning average three-phase load as bus spot-loads, and b) assigning average three-phase impedances as line impedances. Four normally open switches are added (shown with dotted edges in fig. \ref{fig:37bus}). Grid-forming generators are placed at nodes 742, 718 and 710. PV generators of equal rated capacity are added at nodes 702, 705, 707, 709 and 737. There are 22 buses with non-zero load. Total rated capacity of grid-forming and PV generators are 13\% and 29\% of the rated system load respectively. { Such a contrived feeder model has been intentionally chosen to capture potential flexibilities of a much larger network. In practice, a feeder with fewer generators and tie-lines, would have fewer load-generation scenarios and would be faster to solve for.}

Load-generation scenarios were constructed as described next. Data corresponding to hourly solar generation in California from NREL's solar power dataset were used to synthesize five annual generation profiles \cite{solar_data}. The first 50 generators in the dataset were used; every 10 generators were aggregated to obtain one profile. The normalized profiles were then scaled to match the rated capacity of the generators. It is further assumed that the PV generators are set to work at unity power factor, implying that they do not participate in reactive power support. This is without loss of generality since PV generators with reactive power support may be indicated with non-zero entries in the $\boldsymbol{\xi^{g_q}}$ vector. In a similar manner, hourly load profiles were constructed for residential and commercial buildings in California with data available from OpenEI \cite{load_data}. The normalized profiles were scaled such that the 75th percentile of load data coincided with the nominal spot load of the corresponding bus. 
Thus, a total of 8760 scenarios were constructed for a year; denoted as set $\mathcal{S}$.

 \vspace{-0.15in}
\subsection{Chance-Constrained ODNP}

\begin{figure}[t!]
\vspace{-10pt}
    \centering
        \begin{minipage}{0.49\columnwidth}
        \centering
        \includegraphics[trim= 0 0.4in 0 0 clip,width=\linewidth]{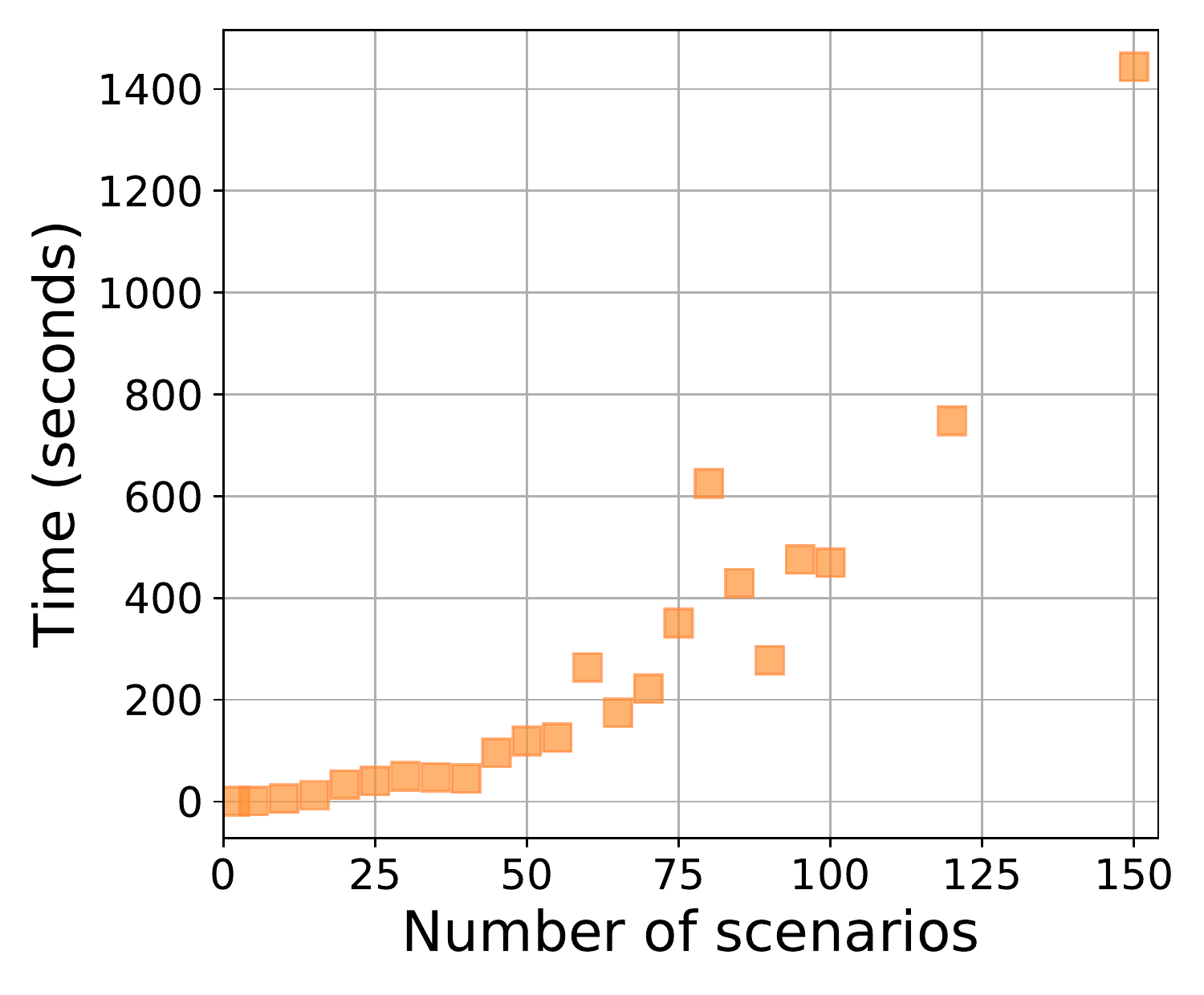}
        \caption{Increase in median computation time with number of scenarios}
        \label{fig:runtime}
        \vspace{-0.05in}
    \end{minipage}\hspace{0.005\columnwidth}
    \begin{minipage}{.49\columnwidth}
        \centering
        \includegraphics[trim= 0 0.4in 0 0 clip, width=\linewidth]{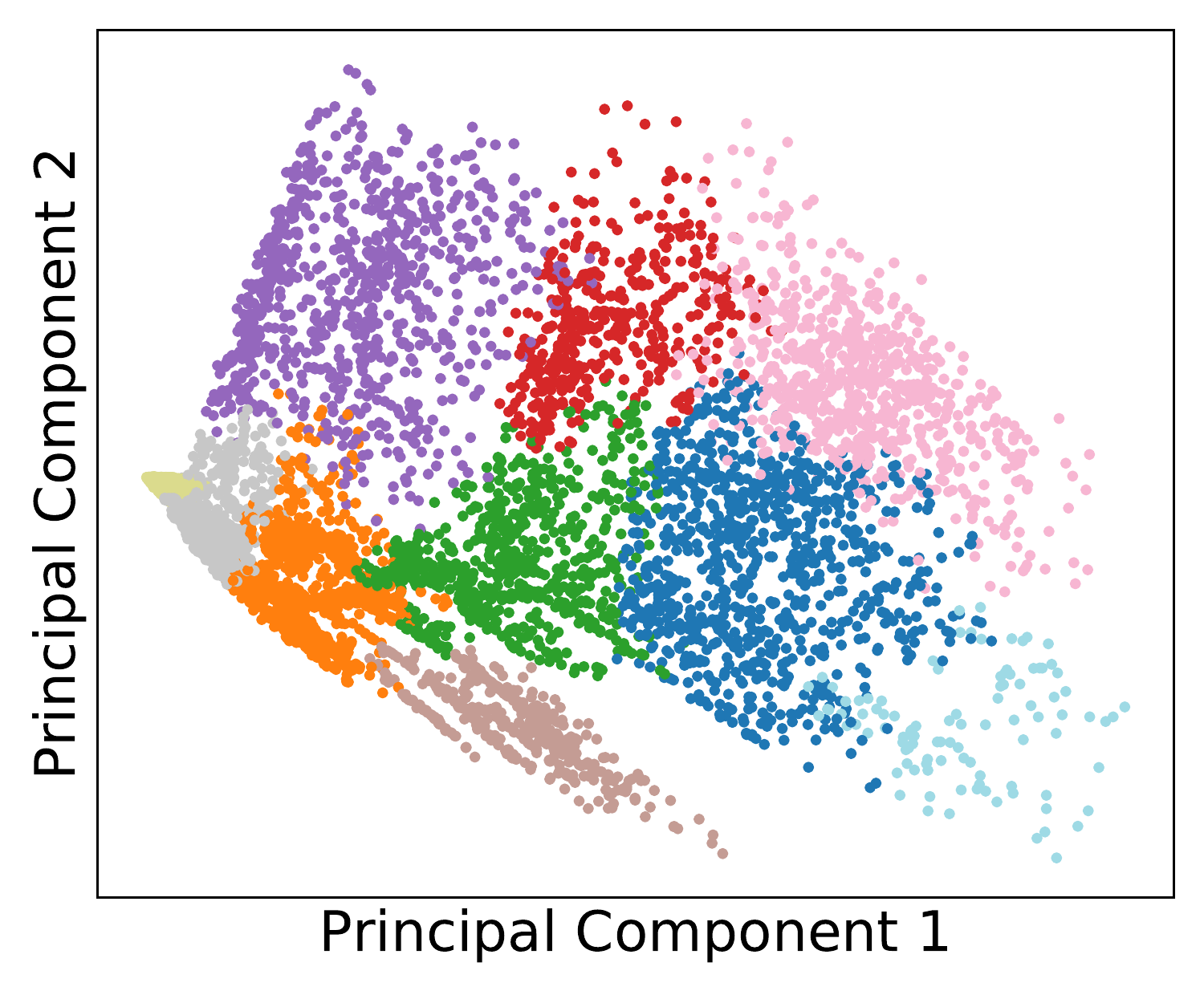}
        \caption{2-D visualization of scenario clustering}
        \label{fig:clusters}
        \vspace{-0.05in}
    \end{minipage}
    \vspace{-0.05in}
    \centering
    \subfloat[Average load served]{\includegraphics[trim=0.1in 0.1in 0.1in 0.14in, clip,width=0.49\columnwidth]{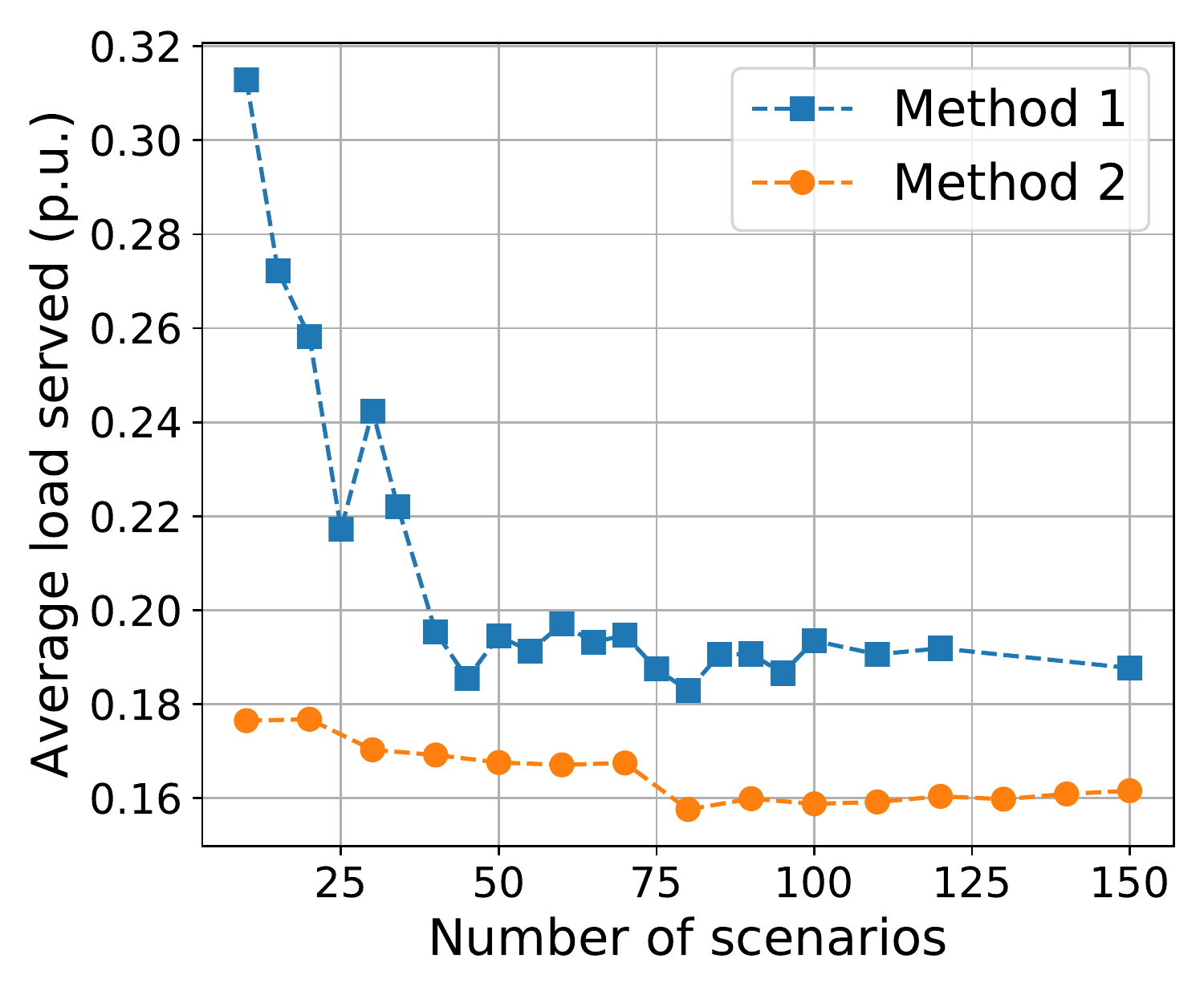}}\hspace{1pt}
    \subfloat[Constraint violation probability]{\includegraphics[trim=0.1in 0.1in 0.1in 0.14in, clip,width=0.49\columnwidth]{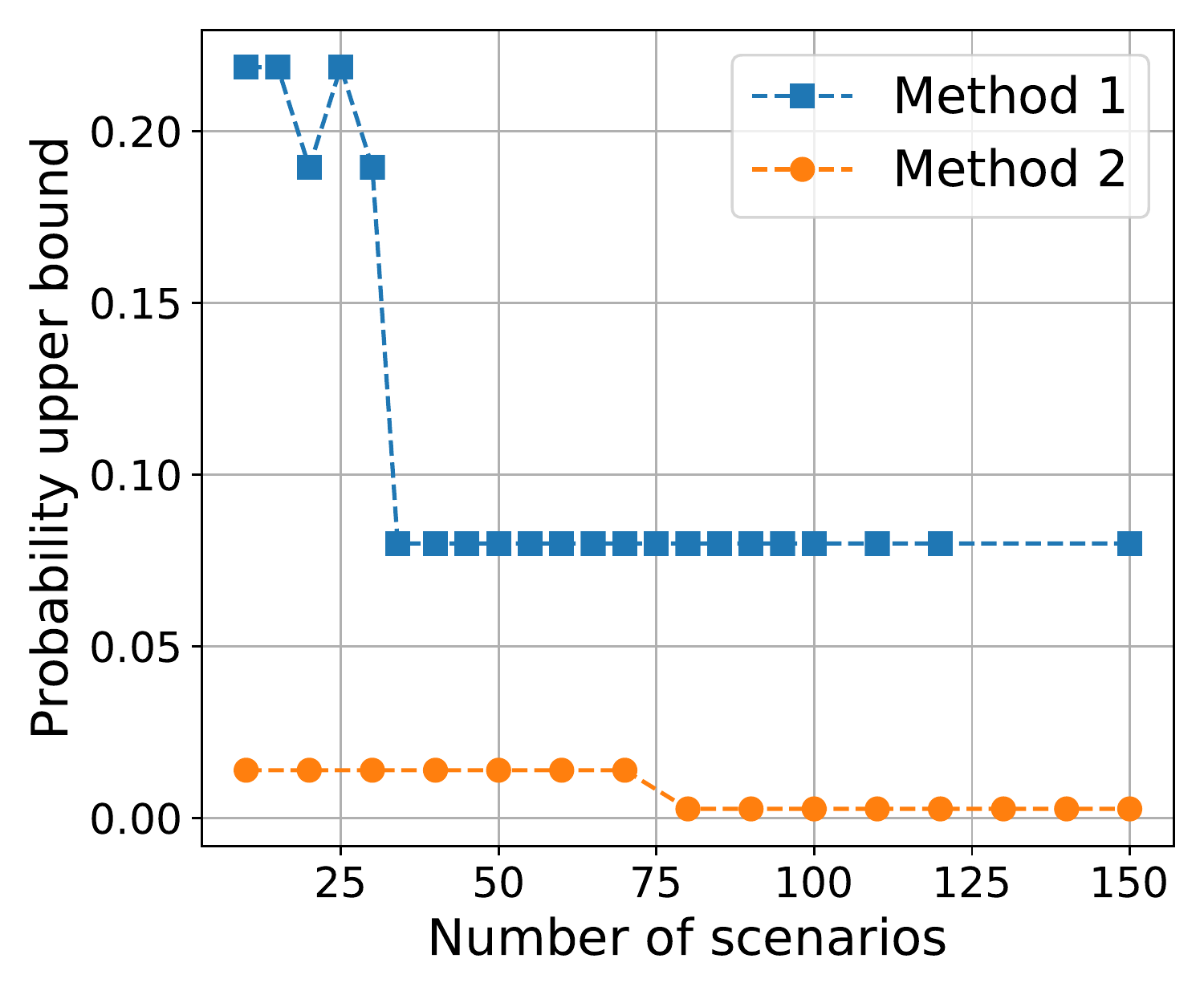}}
    \caption{Variation in performance when number of scenarios is varied from 10 to 150. Proposed cc-ODNP is compared to a clustering based method.}
    \label{fig:oq}
    \vspace{-0.2in}
\end{figure}

As stated previously, the original chance-constrained problem and its SAA counterpart become equivalent in limit as the number of scenarios considered $N$ increases. However, a higher $N$ value also results in high computation time.  This increasing trend is illustrated in fig. \ref{fig:runtime}, the markers show median time for 10 runs conducted over the same scenario sets. For computational tractability, let the SAA problem be solved on a smaller sample set $\mathcal{S'}\subseteq \mathcal{S}$; if $\mathcal{S'}$ is sufficiently representative of $\mathcal{S}$, then the candidate solution obtained will be close to the true solution for the original CCO problem.

Performance of the SAA approach is compared to a clustering based methodology, wherein set $\mathcal{S}$ is divided into clusters and the ODNP task is designed to yield a solution that holds for some representative samples drawn from these clusters. Let us call these two \textit{Method 1} and \textit{Method 2} respectively. 

$\bullet$ \textit{Method 1:} Scenarios are sampled from $\mathcal{S}$ at random with uniform probability and used to solve (ODNP-3).

$\bullet$ \textit{Method 2:} Using principal component analysis followed by hierarchical clustering, set $\mathcal{S}$ is divided into 10 clusters \cite{IDM_tan}. 
The scenario clusters 
are visualized in fig. \ref{fig:clusters}. Once the clusters are determined, equal number of samples are drawn from each cluster at random. Evidently samples can only be drawn in multiples of 10. The ODNP is solved such that the optimal topology is feasible for all selected samples, i.e. $\gamma=0$. 

The performance of the two solution methods is compared in fig. \ref{fig:oq}. For SAA, the value of $\gamma$ used is 0.1. A 95\% confidence lower bound on the objective value is found using the methodology described in Section \ref{sub-sec: validation}. With 50 runs of independently generated sets of 20 scenarios each, and $\gamma=0.7$, this lower bound is determined to be -0.18536. The parameters $M,N''$ and $\gamma$ here were chosen following the recommendations outlined in \cite{lowerbound}. The 95\% confidence upper bound on feasibility of the candidate solution $U_{0.05,1000}(\mathbf{\Bar{x}})$ is estimated using a set of 1000 scenarios. Median computation time for the feasibility checking process was 1.438 seconds. 

Evidently, as more scenarios are considered, both average load served and $U_{0.05,1000}(\mathbf{\Bar{x}})$ decrease. The trends are not strict as additional scenarios can introduce favorable cases with lower cost. Observe that Method 2 yields a more robust solution (low violation probability) in lieu of a higher  objective cost. Method 1 achieves a cost close to the theoretical lower bound while bounding supply-deficiency probability to an acceptable level. Table \ref{tab:optimality} summarizes observations when both methods are run with 100 scenarios. 

\begin{figure}[t!]
\vspace{-0.1in}
    \centering
    \subfloat[Average load served]{\includegraphics[trim=0.1in 0.1in 0.1in 0.14in, clip,width=0.46\columnwidth]{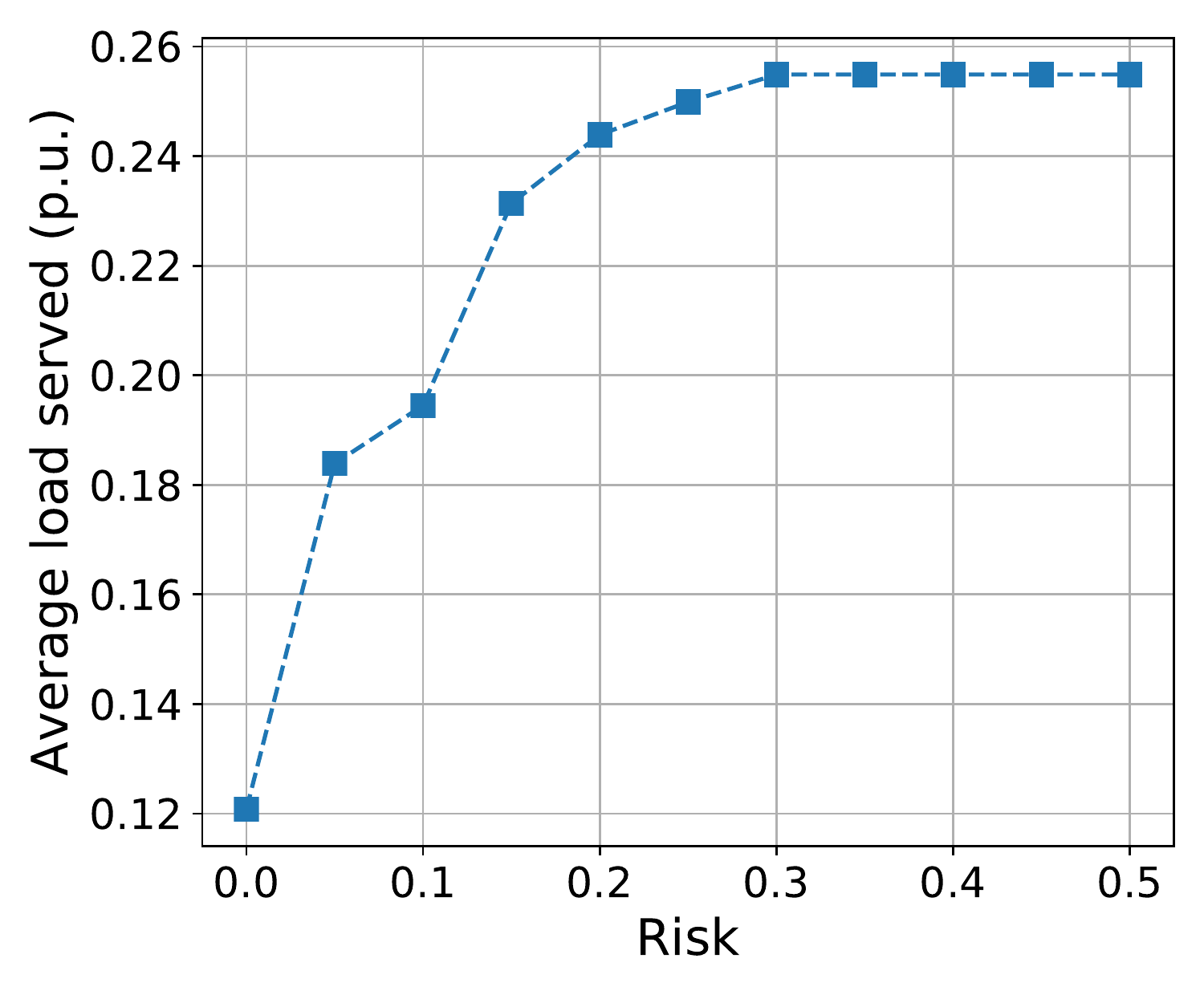}}\hspace{2pt}
    \subfloat[Median time taken for 10 runs]{\includegraphics[trim=0.1in 0.1in 0.14in 0.1in, clip,width=0.46\columnwidth]{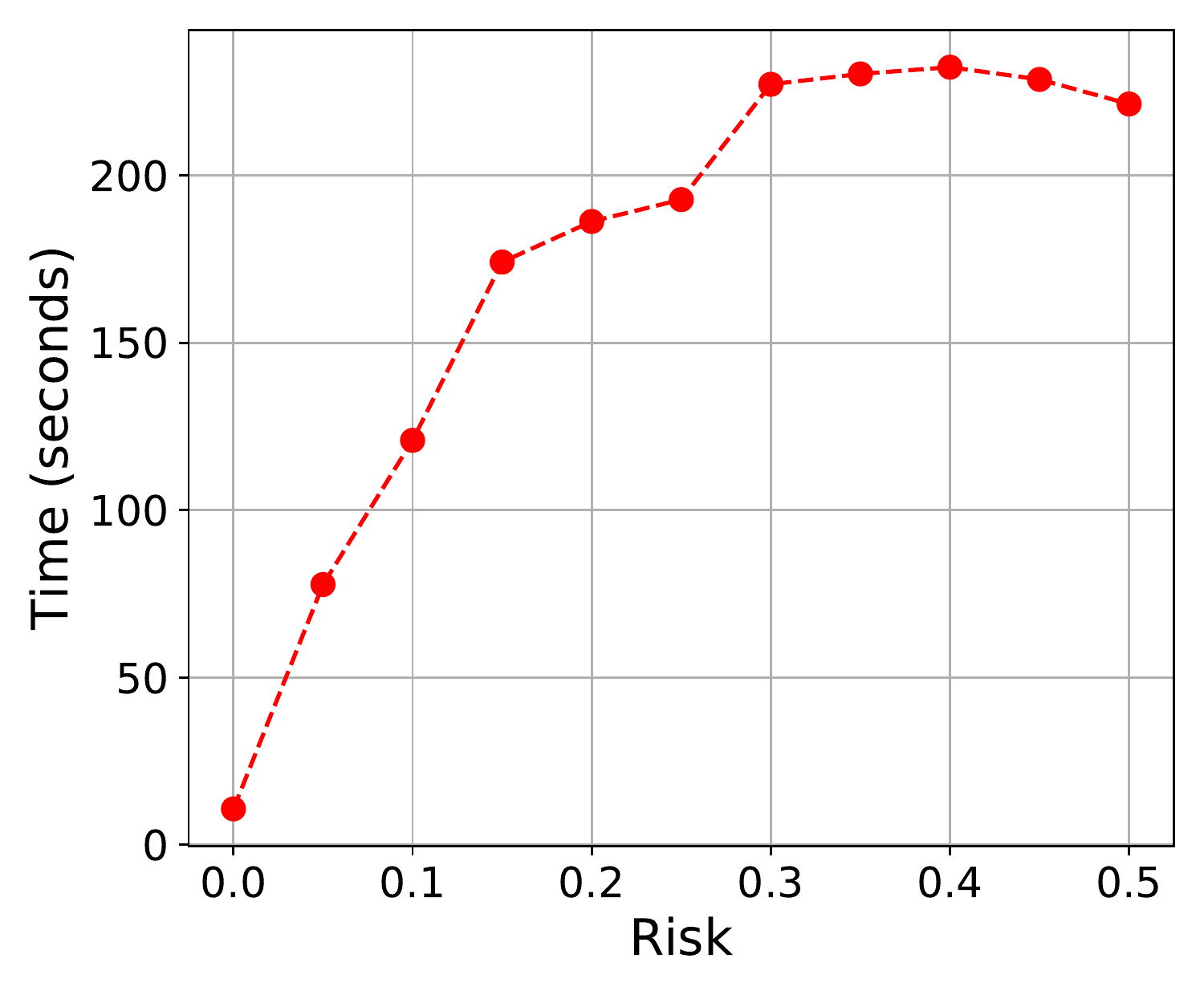}}
    \caption{Variation in cc-ODNP performance with $\gamma$ considering 50 scenarios.}
    \label{fig:risk}
    \vspace{-0.05in}
    \centering
    \subfloat[Average load served]{\includegraphics[trim=0.1in 0.1in 0.1in 0.14in, clip, width=0.46\columnwidth]{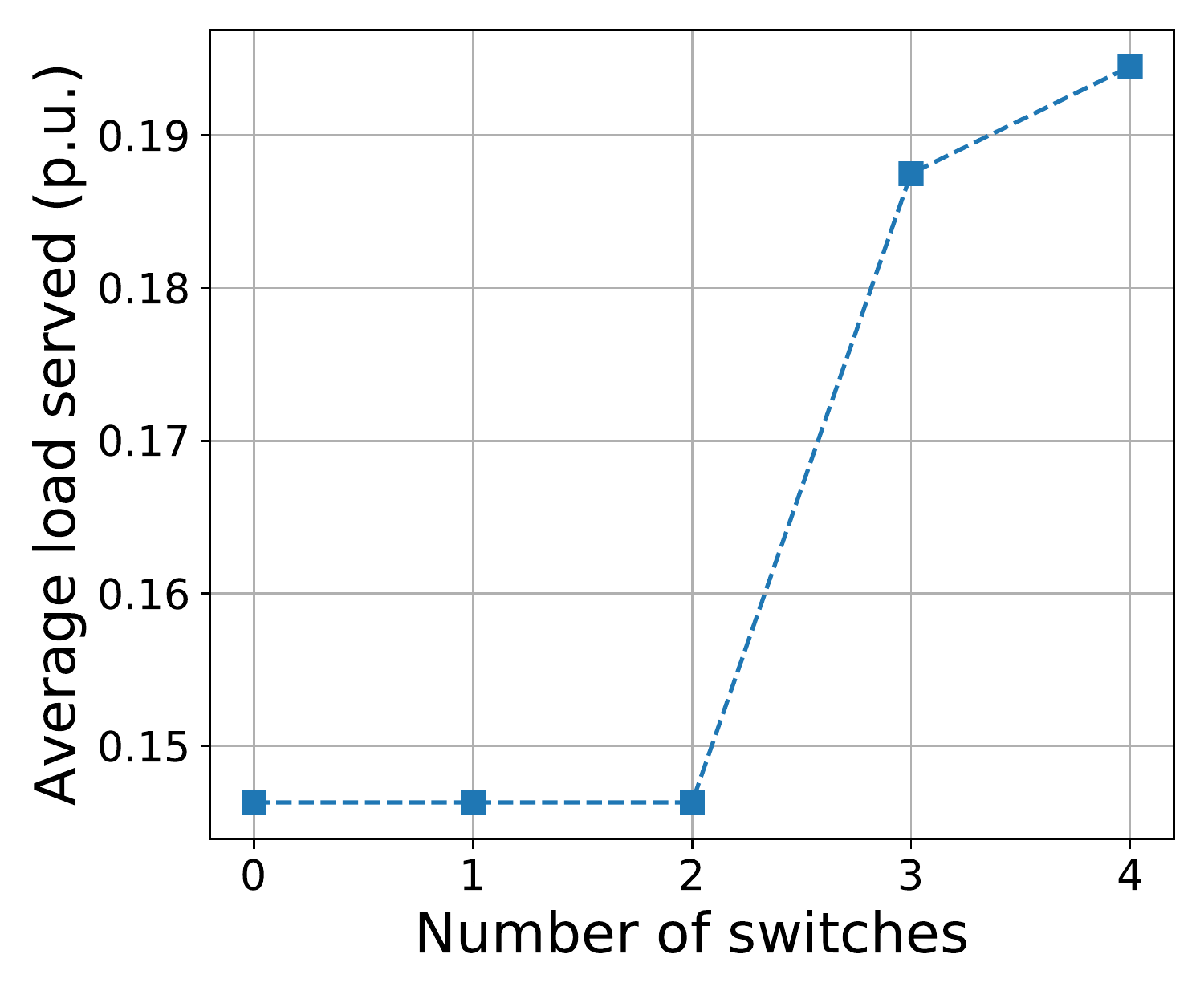}}\hspace{2pt}
    \subfloat[Median time taken for 10 runs]{\includegraphics[trim=0.1in 0.1in 0.1in 0.14in, clip,width=0.46\columnwidth]{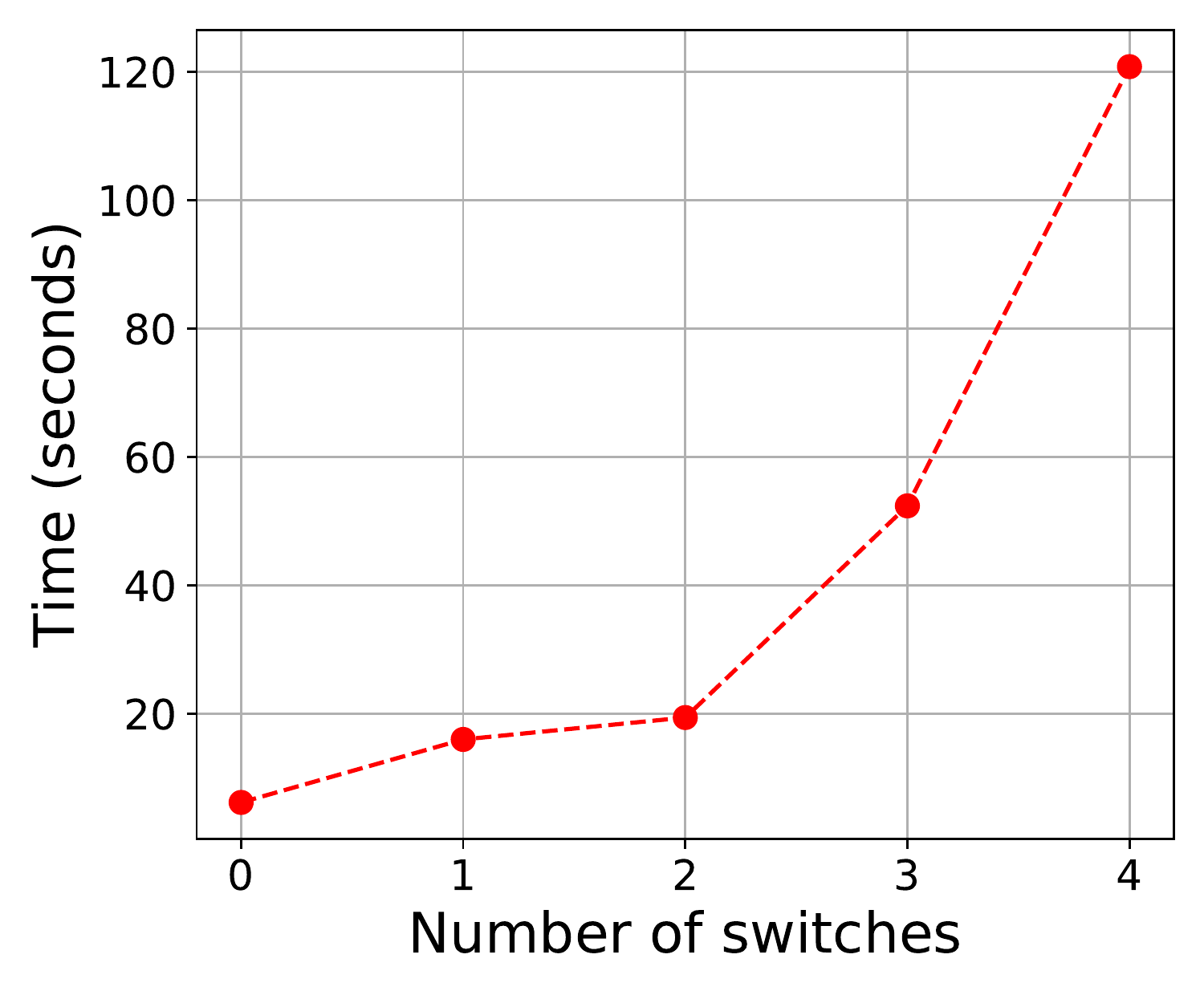}}
    \caption{Variation in performance with number of normally open switches. Number of scenarios considered is 50, $\gamma=0.1$. }
    \vspace{-0.17in}
    \label{fig:switch}
\end{figure}
\begin{table}[t]
    \centering
    \caption{Performance comparison}%
    \begin{tabular}{ccc}
    \hline & \textbf{Method 1} & \textbf{Method 2} \\
    \hline $\mathbf{|f(\Bar{x})-f(x^*)|}$& 0.00814 & 0.34416 \\
    \hline $\mathbf{U_{0.05,1000}(\Bar{x})}$ & 0.08 & 0.0025 \\
    \hline
    \end{tabular}
    \label{tab:optimality}
    \vspace{-0.2in}
    \end{table}
    
    \begin{figure*}[!ht]
    \vspace{-0.05in}
    \centering
    \begin{minipage}{0.49\textwidth}
    \includegraphics[width=0.98 \columnwidth]{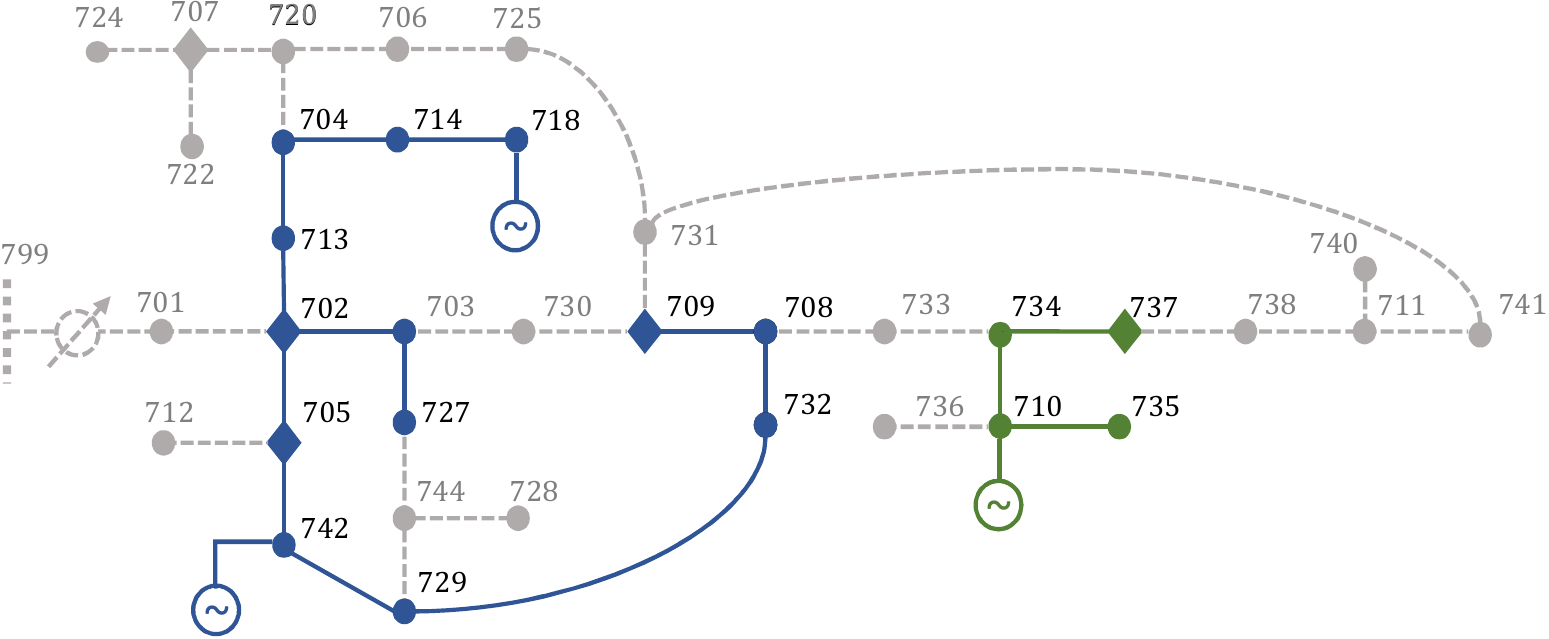}
    \caption{Optimal microgrid topology considering all switches, 100 scenarios and $\gamma=0.1$. Average load served is 0.1935 p.u.. $U_{0.05,1000}(\mathbf{\Bar{x}})= 0.08005$.}
    \label{fig:optimal}
    \end{minipage}\hspace{0.005\textwidth}
\begin{minipage}{0.49\textwidth}
\vspace{-25pt}
    \centering
    \includegraphics[width=0.98\columnwidth]{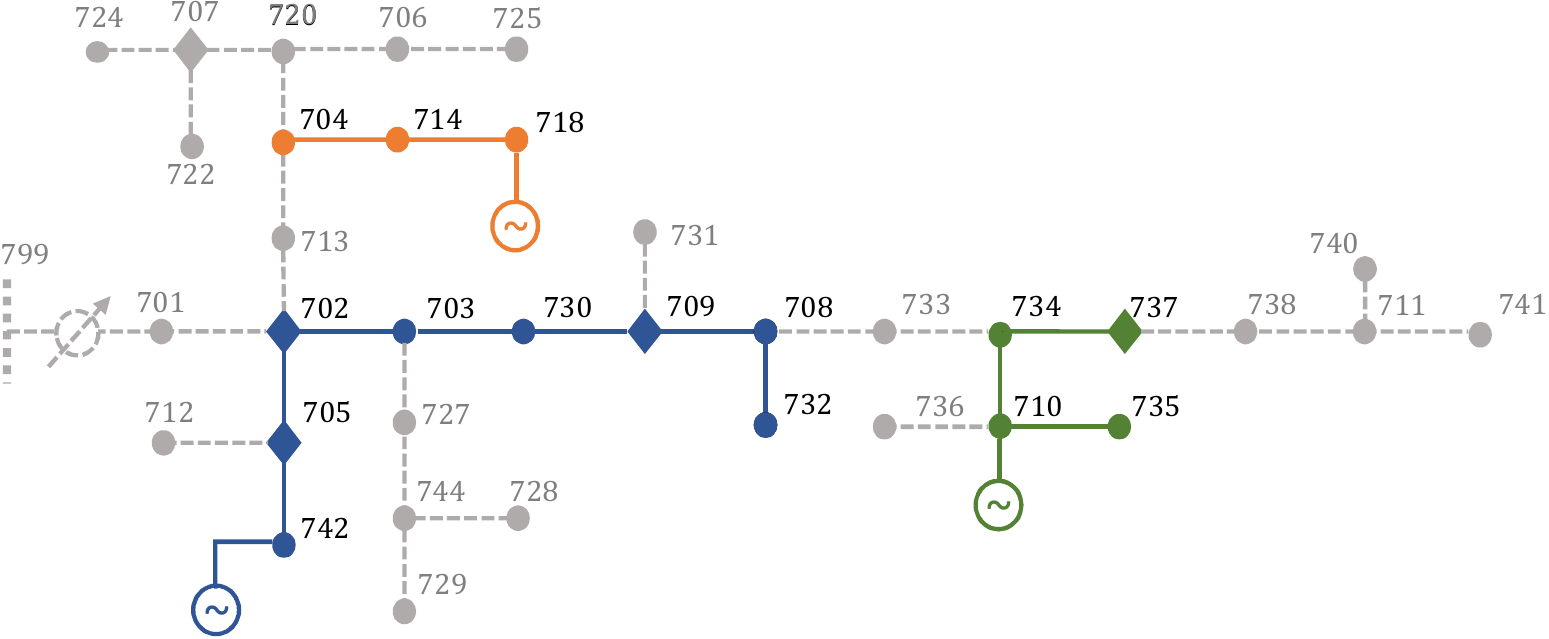}
    \caption{Optimal microgrid topology for the base radial network, 100 scenarios and $\gamma=0.1$. Average load served is 0.1483 p.u.. $U_{0.05,1000}(\mathbf{\Bar{x}})= 0.06465$.}
    \label{fig:0switch}
    \vspace{-0.35in}
    \end{minipage}
    \vspace{-0.22in}
\end{figure*}
\vspace{-0.15in}
\subsection{Choice of risk parameter}
The optimal topology depends highly on the risk parameter. Of course, if a utility has a high risk budget, they may plan to cover a larger amount of loads with the microgrids. The risk appetite may be dictated by a number of factors, like the installed storage capacity and criticality of loads. The intuition of higher load served with higher risk values is experimentally verified and shown in fig. \ref{fig:risk}. It can also be seen that computation time increases with $\gamma$; possibly because for higher values of $\gamma$, the feasibility space that the solver has to search for an optimal solution to ODNP grows in size.
\vspace{-0.1in}
\subsection{Number of Normally Open Switches}
Any topology determination problem is combinatorial in nature, and hence the search space and solution time increases with the number of graph edges. In the ODNP task, network flexibility may be better utilized to serve more load by leveraging normally open switches. However, addition of extra edges introduces additional binary decision variables, thereby increasing solution time. In fig. \ref{fig:switch}, it is shown that as more switches are added to the base radial 37-bus network, ODNP yields higher average load served (i.e. lower objective values), but the computation time goes up. These data points were determined by solving the ODNP problem over 50 scenarios sampled with method 1 and using $\gamma=0.1$. For each of these cases, multiple combinations of normally open switches are possible. However, switches were added one at a time in a random sequence to the base network for illustration. 
\vspace{-0.1in}
\subsection{Microgrid Topology}
Optimal microgrids determined for the base radial network with and without  normally open switches are shown in fig. \ref{fig:optimal} and \ref{fig:0switch} respectively. Microgrid components are indicated in color while external elements are in grey. When all switches are considered, higher load can be served. When only the base radial network is considered, load served by microgrids is lower, and so is the supply-deficiency violation probability.

The determined microgrids do not violate the self-adequacy and power systems constraints for more than $\gamma$ fraction of cases, are radial and contain at least one grid-forming generator. Notice that despite hosting a solar generator, bus 707 is not included in any of the microgrids. This may be because there are no possible ways to connect bus 707 to a grid-forming generator without violating one of the prescribed constraints.

\section{Conclusion}\label{sec: 4_conclusion}\vspace{-0.05in}
The power grid is critical for maintaining essential sectors like healthcare, transportation and emergency services. This has motivated research towards boosting grid resilience. Efficiently planned microgrids can help minimize load interruptions and aid restoration during and after outages. To this end, this work proposes a chance-constrained optimal network partitioning problem and presents a computationally tractable solution methodology. Uncertainty in load and renewable energy generation as well as constraints like maintaining network radiality and availability of grid-forming generators are addressed. Experiments on a modified version of the IEEE 37-bus feeder show that good quality candidate solutions can be found with modest computation cost. Future work will focus on extending the present planning-stage formulations to multi-phase topologies and near real-time applications.   

\bibliographystyle{./bibliography/IEEEtran}
\bibliography{./bibliography/IEEEabrv,./bibliography/microgrid_bib}

\end{document}